\documentclass[12pt]{article}
\usepackage{amssymb}
\usepackage{amsmath}
\usepackage{graphicx}

\def\beq{\begin{eqnarray}}    
\def\eeq{\end{eqnarray}}      

\def\ln{\,\mbox{ln}\,}                  
\def\Tr{\,\mbox{Tr}\,}                  
\def\Box{\square}                       
\def\Log{\,\mbox{Log}\,}                
\def\Det{\,\mbox{Det}\,}                

\def\na{\nabla}                         
\def\pa{\partial}                       

\def\al{\alpha}
\def\be{\beta}

\def\ga{\gamma}
\def\de{\delta}

\def\ka{\kappa}
\def\la{\lambda}

\def\si{\sigma}
\def\om{\omega}
\def\ph{\varphi}

\def\Ga{\Gamma}
\def\De{\Delta}
\def\La{\Lambda}

\def\Om{\Omega}

\begin{document}

\begin{center}

{\large\bf Counting ghosts in the ``ghost-free'' non-local gravity}
\vskip 4mm

{\bf Ilya L. Shapiro$^{a,b,c}$
}
\vskip 4mm

{\sl
(a) \ D\'epartement de Physique Th\'eorique and Center for Astroparticle
Physics, Universit\'e de Gen\`eve,
24 quai Ansermet, CH–1211 Gen\'eve 4, Switzerland
\vskip 2mm

(b) \ Departamento de F\'{\i}sica, ICE, Universidade Federal de Juiz de Fora
\\
Campus Universit\'{a}rio - Juiz de Fora, 36036-330, MG, Brazil
\\
Email address: \ shapiro@fisica.ufjf.br
\vskip 2mm

(c) \ Tomsk State Pedagogical University and Tomsk State University,
Tomsk, Russia
\vskip 3mm
}
\end{center}
\vskip 6mm

\begin{quotation}
\noindent
{\bf Abstract.} \
In the recently proposed non-local theory of quantum gravity one
can avoid massive tensor ghosts at the tree level by introducing
an exponential form factor between the two Ricci tensors. We show
that at the quantum level this theory has an infinite amount of
massive unphysical states, mostly corresponding to complex poles.
\vskip 3mm


\vskip 2mm



\end{quotation}

\section{Introduction}

The general relativity (GR) is a very successful theory of gravity,
but it is perhaps not an ultimate theory. One of the reasons is that
the fourth derivative terms in the action of gravity become necessary
as the UV completion of the theory at the semiclassical level \cite{UtDW}
(see also \cite{birdav,book} for the introduction and \cite{PoImpo} for
a recent pedagogical review). The same
fourth-derivative terms make the theory of quantum gravity (QG)
renormalizable \cite{Stelle-77}. On the other side, fourth
derivatives lead to the massive ghosts in the physical spectrum
of the theory, leading to the violation of of unitarity.

The consistency of the fourth derivative quantum gravity (QG)
can be, in principle, achieved by dealing with the dressed propagator
instead of the classical one \cite{Tomboulis-77,salstr,antomb}. The
main expectation is that the massive ghost poles become unstable
and decay in the far future, such that the asymptotic {\it out}-state
becomes free of ghosts. Unfortunately, the final conclusion concerning
this approach requires a complete non-perturbative knowledge of the
dressed propagator \cite{johnston}, which is unavailable.

Some years ago a completely different approach was proposed by
Tomboulis \cite{Tomboulis-97}. The action of this new theory of
QG has an infinite amount of derivatives. It was discovered
a few years earlier by Tseytlin \cite{Tseytlin-95} that for some
specially tuned form of the non-local action such a theory is
free of ghosts at the tree level while the exponential form
factors remove Newtonian singularity, similar to the much simpler
fourth derivative gravity \cite{Stelle-77}. The approach of
\cite{Tseytlin-95} (see also \cite{Siegel-03}) was to use this
action in the framework of string theory, as an alternative of
the Zweibach ghost-killing transformation of the background
fields \cite{zwei,dere,Tse}. In string theory the
ghost-free non-local action is a kind of a ``final product'',
which is not supposed to gain further quantum
corrections\footnote{However, this does not make it free of
ambiguity related to the third and higher powers of curvature,
similar to the one discussed in \cite{marot}.}.
On the contrary, if one takes the same model as a basis of
quantum gravity \cite{Tomboulis-97}, the following three
important questions should be answered:

\begin{itemize}
\item
First, how to quantize the non-local theory?

\item Second, what is the power counting in a theory with infinite
amount of derivatives?

\item
The third and most difficult question is what happens with the
ghost-free structure of the theory after the quantum corrections
are taken into account?
\end{itemize}


Concerning the first point, the quantization of non-local theories
has been discussed in the literature \cite{Efimov} and is
relatively well-understood. The second issue has been explored in
\cite{Tomboulis-97} and in the more recent publications
\cite{Mode-fin,LMT,BiMa}. The main conclusion is that the power
counting in the non-local theory of \cite{Tomboulis-97} is the same
as in the local higher derivative superrenormalizable QG suggested
earlier in \cite{highderi}.  Moreover, in both cases there is a chance
to make such a QG theory finite. This can be certainly achieved in
the local case \cite{highderi} and very likely in the non-local
one\footnote{In the odd space-time dimensions this can be easily
proved \cite{Mode-fin}.}.

In the present work we will mainly address the third question.
There are strong arguments that at least the most simple example
of the non-local theory suggested in \cite{Tomboulis-97} does not
remain free of ghost-like states at the quantum level. The last
means that the quantum corrections lead to an infinite amount of
the ghost-like states in the dressed propagator. The relation
between ghosts in the classical (naked) and dressed propagators
is almost opposite to what was expected in the fourth-derivative
renormalized theory of QG \cite{Stelle-77,Tomboulis-77,salstr,antomb}.

The paper is organized as follows. In Sect. 2 we present a brief
review of the non-local gravity which is ghost-free at the tree-level.
In Sect. 3 we explain the power counting in the non-local model, here
our consideration mainly follows previous publications
\cite{Tomboulis-97}, \cite{Mode-fin} and \cite{BiMa}, but we try to
make it more transparent, especially by comparing to the local
superrenormalizable QG case \cite{highderi}. Some relevant details
concerning Lagrangian quantization of the non-local theory are
settled in the Appendix. In Sect. 4 it is shown how the ghost-free
structure is violated by quantum corrections to the propagator.
In Sect. 5 we discuss the modified Newtonian limit in the non-local
theory and the possible role played by the ``hidden'' ghosts.
Finally, in the last section we draw our conclusions.

\section{Non-local ghost-free models}

The simplest way to count degrees of freedom in QG is based on
the analysis of the tree-level propagator on the flat background.
In most of the theories this procedure gives the same result as
canonical quantization \cite{GitTyu,book}. In order to explore
the flat-space propagator, the relevant part of the classical
action is at most bilinear in the curvature tensor,
\beq
\label{act}
S = \int d^4 x \sqrt{-g} \,
\Big\{ - \frac{1}{\ka^2} \,R
+ R \, F_1(\Box) \, R
+ R_{\mu\nu} \, F_2 (\Box)\, R^{\mu\nu}
+ R_{\mu\nu\al\be} \, F_3 (\Box) \, R^{\mu\nu\al\be}\Big\}\,.
\eeq
Here $\ka^2=16\pi G$ and $F_{1,2,3}$ are functions of d'Alembertian
operator. The cosmological constant term is set to zero, following
the standard treatment \cite{Stelle-77}.
In order to simplify the action, let us note that the difference
between the term $R_{\mu\nu\al\be} F_3 (\Box) R^{\mu\nu\al\be}$ and
the combination $4R_{\mu\nu} F_3 (\Box) R^{\mu\nu} - R F_3 (\Box) R$
is proportional to the term of the third power in curvature,
${\cal O}(R^3_{\dots})$ (see, e.g., \cite{highderi,NewSing}).
Therefore one can cast the relevant part of
the action (\ref{act}) in the form
\beq
\label{act Phi}
S = \int d^4 x \sqrt{-g} \,
\Big\{ - \frac{1}{\ka^2} \,R
+ \frac12\,C_{\mu\nu\al\be}  \, \Phi(\Box) \, C_{\mu\nu\al\be}
+ \frac12\,R \, \Psi (\Box)\, R\Big\}\,,
\eeq
where $C_{\mu\nu\al\be}$ is the Weyl tensor.
The function $\Psi$ is responsible for the spin-0 part of the
propagator and the function $\Phi$ for the spin-2 part. For the sake
of simplicity, we can mainly concentrate on the spin-2 sector. The
consideration for the $\Psi$-part would be very similar. After the
Fourier transformation, the relevant equation for defining the poles
of the propagator is \cite{Tomboulis-97}
\beq
p^2\,\big[1 \,+\, \ka^2 p^2\Phi(-p^2)\big] &=& 0\,.
\label{spin-2}
\eeq
One can see that there is always a massless pole corresponding
to gravitons. For a constant $\Phi$ there is also a massive pole
corresponding to a spin-2 ghost, which may be also a tachyon.
For a non-constant polynomial function $\Phi$ there are always
ghost-like poles, real or complex \cite{highderi}. However, one
can choose the function $\Phi$ in such a way that there will not
be any other spin-2 pole, except the graviton $p^2=0$. The simplest
example of this sort is \cite{Tseytlin-95}
\beq
1 \,+\, \ka^2 p^2\Phi(-p^2) &=& e^{\al p^2}\,,
\label{exp}
\eeq
where $\,\al\,$ is some constant of the dimension $mass^{-2}$.
One can find other entire functions which have the same
features \cite{Tomboulis-97,Mode-fin}, but for the sake of
simplicity we consider only (\ref{exp}).

Let us remember that the exponential function has two remarkable
properties. The equation $\exp z = 0$ has no real solutions and
only one very peculiar solution
\beq
z = - \infty + i\times 0
\label{inf}
\eeq
on the extended complex plane. At the same time, already the
equation $\exp z = A \neq 0$ has infinitely many complex solutions,
the same is true for
\beq
e^z = A z^2 \log z\,,
\label{exp-brok}
\eeq
which is the typical case for the exponential theory with
logarithmic quantum corrections.
These well-known features of exponential function mean, in our
case, that the absence of massive ghosts in the spin-2 part of
the propagator of the theory (\ref{exp}) is the result of an
absolutely precise tuning of the function $\Phi(-p^2)$.  If this
tuning is violated by the loop corrections, then the ghosts-like
states will emerge in an infinite number. For instance, any
polynomial addition to the exponential function produce infinitely
many complex solutions.

One important note is in order. The expression ``ghosts-like
states'' mean that these states are not exactly the ``classical''
massive ghosts, that means states with positive square of mass
and negative kinetic energy. In the present case there are mostly
complex poles, that means a complex ``square of mass'' and complex
``kinetic energy''. This situation makes the particle interpretation
of these states rather complicated. We postpone the discussion of
this issue until another publication and will call these states
simply ghosts in what follows.

If the theory with more ghosts should be qualified worst, then
the exponential gravity (\ref{exp}) with violated absolute tuning
is worst than the polynomial version of superrenormalizable QG
\cite{highderi} (see also the next section), because the last
has only finite amount of ghosts. So, the main question
concerning the theory of exponential gravity (\ref{exp}) is
whether one can preserve an absolute tuning of (\ref{exp}) at
the quantum level. In the next sections we consider this issue
starting from the strongest effect related to the UV divergences
and related logarithmic running. For comparison, we also present
considerations for the mentioned polynomial model of QG.

\section{Power-counting in local and non-local QG}

Before discussing the dressed propagator
and possible violation of the absolute tuning in (\ref{exp}), let
us shortly review the renormalization properties of the theory
(\ref{act Phi}) and some its natural extensions. A brief survey of
the Lagrangian quantization of the theories such as (\ref{act Phi})
or (\ref{superre}) with some details related to non-local
versions of the theory can be found in the Appendix.

\subsection{Polynomial higher-derivative gravity}

The action of the general superrenormalizable polynomial model can
be written as
\beq
S
&=& S_{EH} \,+\,\int d^4x\sqrt{-g}\,\Big\{
d_1R_{\mu\nu\al\be}^2 + d_2R_{\mu\nu}^2 + d_3R^2 + \, ...
\label{superre}
\\
&+&
c_1 R_{\mu\nu\al\be} \Box^k R^{\mu\nu\al\be}
+ c_2 R_{\mu\nu}\Box^k R^{\mu\nu}
+ c_3 R\Box^k R
\, + \dots + \, b_{1,2,..}R_{...}^{k+2}
\Big\}\,.
\nonumber
\eeq
where the omitted terms and $b_{1,2,..}R_{...}^{k+2}$ denote the set
of all covariant local terms with the derivatives up to
the order $2k+4$. The action includes not only quadratic in curvature
terms, but also generic ${\cal O}(R^3_{...})$ terms, and so on.
$\,d_{1,2,3},\,c_{1,2,3},\,...b_{1,2,\,...}\,$
are arbitrary coefficients.

In order to explore the superficial degree of divergence of the
theory one needs two relations, namely
\beq
D + d \,=\,\sum\limits_{l_{int}}(4-r_l)
\,-\,4n \,+\,4\,+\,\sum\limits_{\nu}K_\nu
\label{dD}
\eeq
for the power counting, and the topological relation
\beq
l_{int} = p + n - 1\,.
\label{topo}
\eeq
In these formulas $l_{int}$ is the number of internal lines with
the inverse power of momenta $r_l$ in the propagator, $n$ is the
number of vertices with $K_\nu$ derivatives and $p$ is the number
of loops.
On the {\it l.h.s.}, $d$ is the number of derivatives acting on the
external lines of a given diagram and $D$ is its superficial degree
of divergence.

In the theory (\ref{superre}) the
most divergent diagrams correspond to the vertices with maximal
number of derivatives, $K_\nu=2k+4$. One can always formulate
the theory (see \cite{Stelle-77,frts-82,highderi} and  Appendix
of the present work) in such a way that $r_l \equiv 2k+4$ for all
fields. Then
it is an easy exercise to combine (\ref{dD}) and (\ref{topo}),
and the result is \cite{highderi}
\beq
d &=& 4\,+\,k(1-p)
\label{sup-pow}
\eeq
for the logarithmically divergent diagrams with $D=0$. The last
relation shows that the versions of QG  with $k \geq 3$ have
only one-loop divergences. This means, the higher order
contributions may be also divergent, but they become finite
after we renormalize the one-loop sub-diagrams. Furthermore,
the possible counterterms may have only four, two and zero
mass dimensions. In other words, only the terms in the first
line of (\ref{superre}) needs to be renormalized. All terms with
derivatives higher than
four are not running. At the same time, the coefficients of these
higher derivative terms define the running of the cosmological
and Newton constants and of the coefficients $d_1$, $d_2$ and $d_3$.

The last two observations which will be used in the rest of the
paper and (as we shall see in what follows) can be applied also
to the exponential gravity, are as follows:

\begin{itemize}
\item
The running of the parameters $\,G$,
$\,\rho_\La\,$ and $\,d_{1,2,3}\,$
is gauge-fixing independent, because the classical equations
of motion have more derivatives than the counterterms. In order
to understand this statement, let us remember that the
gauge-fixing dependence disappears on-shell (see, e.g.,
\cite{FRG-Lavrov} for further references on the subject).
The practical application of this feature to QG was discussed
in \cite{a}.

\item
The $\be$-functions for the Newton constant and the ones of
$\,d_{1,2,3}\,$ are given by bi-linear and linear combinations
of the coefficients of the ${\cal O}(R^3_{...})$ and
${\cal O}(R^4_{...})$ terms in the action (\ref{superre}).
Therefore, one can provide to these $\be$-functions any
desirable values by changing the corresponding
coefficients\footnote{
Of course, this does not mean that the explicit
derivation of these $\be$-functions would not be interesting.
Since the potential result is a possibility to obtain exact
$\be$-functions in some model of quantum gravity, this calculation
would worth the requested hard work anyway. From the physical side,
different choices of these coefficients may correspond to different
physical properties of the theory, so such a calculation would be
quite relevant.}.
The remarkable exception is the $\be$-function for the cosmological
constant derived in \cite{highderi}. This unique $\be$-function is
completely defined by the coefficients $c_{1,2,3}$ in (\ref{superre}).
\end{itemize}

\subsection{Exponential gravity}

In the exponential gravity theory the power counting
formula (\ref{dD}) has no much sense, because it leads to an
indefinite output of the $\infty-\infty$ type. At the same
time the topological relation (\ref{topo}) is working well
and shows that the theory is superrenormalizable
\cite{Tomboulis-97,Mode-fin,BiMa}. Let us consider
this point.

Each propagator gives contribution of
infinite negative powers of momenta, let us call it
${\cal I}$. With the vertices the situation is more complex,
because there are vertices with different powers of momenta.
Without loss of generality one can consider only the
diagrams with maximal divergence, when each vertex gives
contribution $-{\cal I}$. It is important that the two
symbols ${\cal I}$ and $-{\cal I}$ correspond to the same
power of infinity, for otherwise the relation should become
more complicated\footnote{This means, in particular, that
the value of $\al$ in the Eq. (\ref{exp}) must be identical
for both functions $\Phi(\Box)$ and $\Psi(\Box)$ in
(\ref{act Phi}), for otherwise the theory would be badly
non-renormalizable.}.
Then it is clear that
the diagram with more internal lines than vertices will be
automatically convergent and the diagram with more vertices
than internal lines will be strongly divergent. The relation
(\ref{topo}) tells us that the difference is
$\,l_{int} - n = p - 1$. This means that only the one-loop
diagrams with $p=1$ can be divergent. At the same time, the
presence of the exponential form factor does not change the
degree of divergence of the one-loop diagrams.

The power counting in the exponential gravity is performed by
the topological relation (\ref{topo}), without the formula
(\ref{dD}). Nevertheless, the result is exactly the same as
in the polynomial theory (\ref{superre}) for $k \geq 3$.
Namely, the divergences show up only at the
one-loop level, and the counterterms have zero, two and
four derivatives of the metric only. In other words, the
possible counterterms have the form
\beq
\De S
&=& \int d^4x\sqrt{-g}\,\big\{
a_1R_{\mu\nu\al\be}^2 + a_2R_{\mu\nu}^2 + a_3R^2
 + a_4 \Box R + a_5R + a_6\big\}\,.
\label{DeltaS}
\eeq
The divergent coefficients $\,a_{1,2,...,6}\,$ are at most
${\cal O}(1/(n-4))$ in dimensional regularization.

Similar to the polynomial case, there is a chance to specially tune
the $\,{\cal O}(R^2_...)$, $\,{\cal O}(R^3_...)\,$ and
$\,{\cal O}(R^4_...)$-terms in the action (\ref{superre}) such that
the divergent coefficients $\,a_{1,2,...,6}\,$ in (\ref{DeltaS})
can be adjusted to have desirable values. In the case of
exponential QG one should try to provide these divergent
coefficients to become zero, because the possible running would
violate an absolute tuning requested by the ghost-free structure
of the exponential gravity. In case of the logarithmic divergences
of the form (\ref{DeltaS}), the equation for the poles of the
propagator has the form (\ref{exp-brok}) with $z=p^2$ and
$A=4a_1+a_2$ for the spin-2 sector of the propagator. As we already
know, if such corrections take place, then the dressed propagator
has infinitely many ghost-like states. So, if we intend to keep the
ghost-free structure at the quantum level, the first thing to do
is to require that the theory should be finite. We shall discuss
this subject further in the next section.

\section{Quantum corrections and dressed propagator}

As we already know, the power counting in the theory with
exponential form factors (\ref{exp}) is exactly the same as in the
polynomial theory (\ref{superre}) with $k \geq 3$. One can see this
similarity in the following way. Imagine we replace the exponential
function in (\ref{exp}) by the partial sum of its Taylor expansion,
\beq
P_N(\al p^2) &=& \sum\limits_{l=0}^{N}
\frac{\big(\al p^2\big)^l}{l!}\,.
\label{series}
\eeq
For a sufficiently large $N$ the theory will be superrenormalizable,
exactly as in the exponential case. At the same time, there will be
$N$ roots of the polynomial $P_N(\al p^2)$ at the complex plane.
The number of these roots is growing with larger $N$. When
$N\to\infty$ the power counting remains the same. At the same time,
the number of poles becomes infinite, so the theory gains an infinite
amount of ghost-like poles, most of them complex. However, it happens
that all these poles converge to the very special infinite point
(\ref{inf}). Then the theory is ghost-free at the tree-level, but
there is a danger that the absolute tuning may be broken by the
one-loop corrections, which require very special attention.

The one-loop effective action in the theory
(\ref{act Phi}) is given by the expression \cite{frts-82}
(see Appendix for details)
\beq
\Ga_{div}^{(1)} &=&
\frac{i}{2}\,\Tr \Log {\hat H}
\,-\,i\Tr \Log {\hat H}_{ghost}
\,+\,\frac{i}{2}\,\Tr \Log {\hat Y}\,.
\label{one}
\eeq
The last two terms are contributions of ghost and weight operators.
Both of them have standard form plus some part related to the term
$\Tr \Box$.
This term is discussed in the Appendix, where we argue that its
contribution has the form (\ref{DeltaS}) with fixed coefficients.
Therefore, the main question is whether the finiteness can be
provided by changing the action $S$ in such a way that the operator
\beq
{\hat H}  &=&
\frac{1}{2\,\sqrt{-g}}\,\frac{\de^2 S}{\de g_{\al\be}\de g_{\rho\si}}
\label{hess}
\eeq
provides a cancellation of the first term in the formula (\ref{one})
with the divergences coming from the last two terms of the same
expression.

It is obvious that one can not achieve this goal by using the
original action (\ref{act Phi}), because both functions
$\Psi$ and $\Phi$ are proportional to the same expression
\beq
\Psi = c_1\,e^{-\al\Box}\,,\qquad
\Phi = c_2\,e^{-\al\Box}\,.
\label{PsiPhi}
\eeq
It is easy to see that by changing the coefficients $c_1$ and
$c_2$ one modify only the cosmological constant - type counterterm,
and not the fourth-derivative ones, which are relevant for Eq.
(\ref{exp-brok}). Therefore, in order to provide finiteness one
has to generalize the action (\ref{act Phi}). As it was discussed
for the polynomial QG, this can be done by
adding $\,{\cal O}(R^3_{...})$- and $\,{\cal O}(R^4_{...})$-type
terms. The explicit calculation in this theory would be quite
difficult and also there is no real need to make it. Let us instead
present a general evaluation of the possible effect of the
${\cal O}(R^3_{...})$-type terms. The general form of the terms
with a minimal possible non-local insertion is
\beq
\frac{1}{M^2}\,\int \sqrt{-g}\,
R_{\dots}
R_{\dots}
\,e^{-\al\Box}\,
R_{\dots}\,,
\label{R3}
\eeq
where $M$ is a new massive parameter.
Then the operator (\ref{hess}) will have a general non-minimal
structure (after an appropriate gauge-fixing)
\beq
{\hat H}  &\propto&
{\hat \Box} \,+\, M^{-2}\,\hat{\cal D}^{\mu\nu} \na_\mu\na_\nu
\,+\, {\hat \Pi}\,,
\label{hess-nm}
\eeq
where all the operators act in the space of quantum metrics, and
$\hat{\cal D}^{\mu\nu}$ and ${\hat \Pi}$ are proportional to the
curvature tensor. The contribution of $\Tr\Log {\hat H}$ for the
operators of the form (\ref{hess-nm}) is known, in particular it
was elaborated recently in \cite{CPTL-Tib} by means of the
generalized Schwinger-DeWitt technique \cite{bavi-85}. One meets
the one-loop divergences which are given by an infinite series in
curvatures $\hat{\cal D}^{\mu\nu}$ and ${\hat \Pi}$, and the
(super)renormalizability of the theory is completely broken, so
we have to look for some generalization.

Another
possibility is to modify the expression (\ref{R3}) by introducing
further non-localities. The possible solution is to consider
the non-local terms of the general form
\beq
\int \sqrt{-g}\,
R_{\dots}
\frac{1}{\Box}
R_{\dots}
\,e^{-\al\Box}\,
R_{\dots}\,.
\label{R3-N}
\eeq
The one-loop divergences in the theories of similar type were already
considered in the literature \cite{Ishinose}.
Let us note that the expression (\ref{R3-N}) still leaves us a lot of
freedom in the choice of the action, because of the numerous possible
tensor structures and corresponding coefficients.  Since the number
of the possible tensor structures in the operator ${\hat H}$ is
restricted, there is a good chance to meet such a combination of
terms in (\ref{R3-N}) which would lead to the operator
\beq
{\hat H}  &\propto&
{\hat \Box}^2 \,+\, \hat{V}^{\mu\nu} \na_\mu\na_\nu
\,+\, {\hat U}\,,
\label{hess-N}
\eeq
plus some contribution of the operator ${\hat \Box}^{-1}$ which can
be factorized out in a standard way (see, e.g., Chapter 9 of
\cite{book}). In the expression (\ref{hess-N}) one still has the
freedom to choose the operator $\hat{V}^{\mu\nu}$, which is
proportional to the curvature tensor. As a result, it is possible
to manipulate the divergent part of effective action (\ref{DeltaS})
and to provide the desirable pre-fixed values for the coefficients
$a_{1,2\dots 6}$. In particular, there is a chance to obtain a
finite QG in this way.

The situation may be even more simple if  we include the
${\cal O}(R^4_{...})$-type terms with an additional $\,\Box^{-2}$
insertion. In this case the relevant operator will have the form
\beq
{\hat H}  &\propto&
{\hat \Box}^3 \,+\, \hat{V}^{\mu\nu\al\be} \na_\mu\na_\nu \na_\al\na_\be
 \,+\, \hat{U}^{\mu\nu} \na_\mu\na_\nu
\,+\, {\hat W}\,,
\label{hess-N4}
\eeq
similar to the one we dealt with in \cite{highderi}. The operator
$\hat{U}^{\mu\nu}$ will be linearly proportional to the
coefficients of the ${\cal O}(R^4_{...})$-type terms. On the
other hand, linear dependence will also take place between
$\hat{U}^{\mu\nu}$ and the fourth-derivative terms in (\ref{DeltaS}).
Therefore, there are pretty good chances to provide finiteness in the
exponential QG theory by means of a special choice of the coefficients
of the ${\cal O}(R^4_{...})$-type terms with an appropriate non-local
insertion.

Indeed, the possibility to have a finite theory in the non-local
case is not so certain as in the polynomial QG (\ref{superre}).
In case of the non-finite theory the ghost-free structure will
be certainly violated. So, let us be generous to the exponential
QG and simply assume that the non-local theory can be made finite
in the way we described above. As we shall see right now, this
is still not sufficient to prevent the theory from the ghost-like
states. The consistent theory of QG should include quantization
of matter fields, not only the metric. The matters fields of the
spin-0, spin-$1/2$ and spin-$1$ contribute to the divergences in
the form of Eq. (\ref{DeltaS}) \cite{birdav,book}. For the
illustration purpose, let us reproduce the complete form factors
of the one-loop quantum corrections to the $\Phi$-function in Eq.
(\ref{act Phi}), derived in \cite{apco,fervi} for massive scalar
and fermion fields,
\beq
{\bar \Ga}^{(1)}_{scal}
&=&
\frac{1}{32\pi^2}\,\int d^4x \sqrt{-g}\,
C_{\rho\si\al\be} \Big[\frac{1}{60\,(4-n)}
+ \frac{1}{120}\,\ln \Big(\frac{4\pi \mu^2}{m^2}\Big)
+ \frac12\,k^s_W(a)
\Big] C^{\rho\si\al\be}
\label{final}
\eeq
and
\beq
{\bar \Ga}^{(1)}_{ferm}
&=&
\frac{1}{32\pi^2}\,\int d^4x \sqrt{-g}\,
C_{\rho\si\al\be} \,\Big[\,\frac{1}{10\,(4-n)}
+ \frac{1}{20}\ln \Big(\frac{4\pi \mu^2}{m_f^2}\Big)+\frac12\,k^f_W(a)
\Big] C^{\rho\si\al\be}\,,
\label{final-f}
\eeq
where
\beq
k_W(a)
&=&
\frac{240A + 20a^2 + 3a^4}{450\,a^4}\,,
\label{ff-scal}
\\
k^f_W(a)
&=&
\frac{300Aa^2-480\,A
\,-\, 40 a^2 \,+\, 19\,a^4}{225\,a^4}
\label{ff-ferm}
\eeq
and we used notations
\beq
A &=&
1-\frac{1}{a}\ln \frac{1+a/2}{1-a/2}
\quad {\rm and} \quad
a^2 \,=\,
\frac{4\Box^2}{\Box^2-4m^2}\,.
\label{A}
\eeq
The contributions to the $\Psi$-function are qualitatively similar
\cite{fervi}, but we do not reproduce them here for the sake of brevity.

The first observation is that the divergences in the Weyl-squared
sector have the same sign independent on whether we take scalar,
fermion, massless or massive vector. This well-known feature of
the divergences \cite{birdav} means that no cancelation of the
overall contribution to the $\Phi$-function due to supersymmetry
is possible. Therefore, if there is no cancelation with the
divergences coming from the QG sector, the expression (\ref{exp})
gains $\,p^4\log p^2\,$ contribution due to the matter fields loops
and this is certainly sufficient to have infinitely many ghost-like
excitations at the quantum level.

The second important point is that, even if the cancelation of
the divergences really takes place, it is not sufficient to
preserve the ghost-free structure of the theory even at the
one-loop level. The reason is that
both expressions (\ref{ff-scal}) and (\ref{ff-ferm}) have
an infinite sets of sub-logarithmic contributions, and those
can not be canceled by the QG part. The situation may be, in
principle, different in a strictly massless theory of matter,
when the cancelation in the leading-log part may be sufficient.

Furthermore, even if one can adjust the QG contribution to
make the theory completely free of divergences in the
presence of matter fields, this would be true only at the
one-loop level. Let us remember that the exponential QG has
only one-loop divergences, but for the matter loops this is
not so. Starting from the second loop, matter fields produce
the form factors with higher powers of $\log \big(\Box/\mu^2\big)$
in the UV. Then the compensation seems to be completely
impossible.  Let us note that the $\beta$-functions in the
matter sector are not affected by QG in all superrenormalizable
models, as it was explained in Sect. 5 of \cite{highderi}.

One can naturally expect that the same breaking will take place in
the case of pure QG without matter. Indeed, any quantum theory,
including (\ref{act Phi}) may produce sub-logarithmic
contributions in the dressed propagator, and then an absolute
fine-tuning leading to the ghost-free structure will be violated.
Unfortunately, the explicit results concerning sub-logarithmic
contributions in QG are not available, but there can not be much
doubts about their existence. And the last is sufficient for
violating an absolute tuning leading to the ghost-free structure
of exponential gravity (\ref{exp}). So we have to conclude that
the exponential QG has an infinite set of massive ghost-like
states at the quantum level.

\section{Note concerning Newtonian singularity}

There is a very interesting and simple relation between the
renormalizability of the QG theory and the absence of Newtonian
singularity at the classical level. This relation was first
noted by Stelle in \cite{Stelle-77}, the main formula for the
modified Newtonian potential is
\beq
\ph (r) = - GM \left[\frac{1}{r}
- \frac{4}{3} \frac{e^{-m_{(2)} r}}{r}
+ \frac{1}{3} \frac{e^{-m_{(0)} r}}{r} \right]\,.
\label{eq9}
\eeq
Here $m_{(2)}$ and $m_{(0)}$ are masses of the spin-2 ghost
and spin-0 massive particle which are present in the spectrum
of the fourth-derivative gravity. It is easy to see that near
the origin $r=0$ the contribution of these two massive degrees
of freedom exactly cancels the one of the graviton, such that
the limit of the modified Newtonian potential $\ph(r)$ at
$r \to 0$ is free of singularity. In our recent work
\cite{NewSing} it was shown that the same cancelation takes
place in the more general theory of gravity with the action
(\ref{superre}), if additional degrees of freedom in this
theory correspond to the non-degenerate real massive
poles\footnote{The spin-0 contribution was elaborated much
earlier in \cite{HJS}}. For a while, there is no proof that
the same effect takes place in the case of complex massive
poles.

It is remarkable that the non-singular modified Newtonian limit
takes place also in the exponential gravity theory (\ref{exp}).
This was originally found by Tseytlin in \cite{Tseytlin-95}.
Recently, exactly the same non-singular solution has been
rediscovered in the papers \cite{Mode-fin,BGKM,BiMa}. An unfortunate
detail apparently related to the Ref. \cite{BGKM} is the evaluation
of our preprint \cite{NewSing} which was done by Biswas et al in
\cite{BiMa}. On page 3 of their manuscript authors say ``{\it It
was not until recently though, that concrete criteria for any
covariant gravitational theory (including infinite-derivative
theories) to be free from ghosts and tachyons around the Minkowski
vacuum was obtained by Biswas,
Gerwick, Koivisto and Mazumdar (BGKM) [23,24], see also [25]
for a recent re-derivation of the same results using auxiliary
field methods.}'' \footnote{Here {\it [23,24]} correspond to
\cite{BGKM} and \cite{BGKMsemG} and {\it [25]} to the citation
\cite{NewSing} of the present work. Let me stress that our
paper \cite{NewSing} is devoted to the Newtonian singularity
in the polynomial theory (\ref{superre}) and hence we did not
repeat the result of \cite{Tseytlin-95} and consequent works
such as \cite{Siegel-03} and \cite{BGKM}, which calculated the
modified Newtonian limit in the exponential theory case.}.

I believe that the evaluation of our work which was done in
\cite{BiMa} is not correct for at least two reasons. First, in
our Ref. \cite{NewSing} there is no use of auxiliary fields.
Second, we did not explore or discuss ``concrete criteria
for any covariant gravitational theory (including
infinite-derivative theories) to be free from ghosts and
tachyons around the Minkowski vacuum'' and moreover we even
did not deal with the ``infinite-derivative theories'' in
\cite{NewSing}. As it was already mentioned above, our work
\cite{NewSing} is about Newtonian singularity in the polynomial
theory (\ref{superre}), so it is not easy to understand what
the observation of \cite{BiMa} actually means.

Coming back to the relation between (super)renormalizability of
QG and the absence of Newtonian singularity, perhaps the most
intriguing aspect is that the theory without real poles can be
free of singularities. One can note that the polynomial theory
with a form factor given by the partial sums of the exponential
function (\ref{series}) has no real poles in the propagator.
From the other side, the limit at $N\to \infty$ is free of
Newtonian singularity, according to \cite{Tseytlin-95} and to
the consequent publications on the subject \cite{Siegel-03,BGKM}.
Let us remember that the polynomials with growing $N$ have
growing amount of massive poles. One most natural physical
interpretation is as follows. These poles are organized in such
a way that they lead to the cancelation of singularity in a way
similar to (\ref{eq9}) and to the more complicated relations
discussed in \cite{NewSing}. Then the cancelation of singularity
in the exponential case of \cite{Tseytlin-95} is nothing
else but the same effect coming from an infinite amount of
the ``hidden'' ghosts. This conjecture is something
interesting to verify, in our opinion it would give better
understanding of the relation between local and non-local
models of QG.

\section{Conclusions and discussions}

The main conflict of QG is between renormalizability and unitarity.
In order to have renormalizable or superrenormalizable QG, one has
to include higher derivatives into the starting action. Higher
derivatives lead to ghosts and/or tachyons and excluding these
unphysical states from the spectrum produce violation of unitarity.
Since higher derivatives emerge already at the semiclassical level,
apparently there is no way to avoid them, so the question is
how to deal with the massive ghosts.

Some qualitatively new approach to the problem of QG was suggested
in \cite{Tomboulis-97} and recently elaborated further in
\cite{Mode-fin,LMT} and \cite{BiMa}. This new approach assumes
that the starting theory is chosen in such a way that the higher
derivative theory is free of ghosts from the very beginning.
Such a choice implies that the theory must be non-local, in
a way explored earlier by Efimov et al \cite{Efimov}.

We have shown that in the case of exponential QG the usual power
counting evaluation must be modified and the main role is played
by the topological relation between the number of vertices and
internal lines. After all, the renormalization properties of the
non-local theory of \cite{Tomboulis-97} are very similar to the
ones of the superrenormalizable QG, introduced earlier in
\cite{highderi}. In particular, this means that the $\be$-functions
of the matter fields are not affected by QG and that the
$\be$-functions in the gravitational sector can vanish,
rending the theory finite.

While the classical theory of exponential gravity is ghost-free,
the quantum corrections may easily lead to the dressed propagator
which has infinitely many complex ghost-like poles. We have shown
that this scenario is unavoidable if the theory is not finite.
The finiteness in such a theory is possible by tuning the
${\cal O}(R^3_...)$ and ${\cal O}(R^4_...)$-type non-local terms
in the action. This may guarantee the absence of the strongest
logarithmic corrections and the result can be extended even to
the one-loop theory with quantum matter included. However, at
higher loops this tuning breaks down. On the other hand, even
the one-loop contributions of massive matter fields have
well-established sub-logarithmic contributions, which can
not be cancelled in the exponential QG model. The final conclusion
is that an infinite amount of unphysical complex poles emerge in
the theory at the quantum level.

In our opinion, an improved understanding of the role of ghosts is
one of the most relevant issues for QG. In particular, the main
lesson which one should learn from the comparison of polynomial
and exponential models of QG is the importance of the models with
complex poles, which were not covered by the consideration of
\cite{highderi}. It would be
interesting to treat this case in details in both quantum field
theory framework and in the more phenomenological way proposed
recently in \cite{HD-Stab}.

\section*{Acknowledgements}
Author is grateful to CNPq, FAPEMIG and ICTP for partial support
of his work and to the D\'epartement de Physique Th\'eorique and
Center for Astroparticle Physics at the Universit\'e de Gen\`eve
for partial support and kind hospitality during his sabbatical
period. I am also grateful to Theoretisch-Physikalisches-Institut
of the Friedrich-Schiller-Universit${\ddot {\rm a}}$t in Jena for the
warm hospitality during the time when the first version of this
paper was completed. The observations of anonymous referee were 
very useful in improving presentation and I would like to 
express gratitude to his/her contribution.  

\section*{Appendix. Brief review of Lagrangian quantization}

Let us consider the Faddeev-Popov procedure for the theory
of QG based on the action (\ref{act Phi}) with possible
additional terms of the ${\cal O}(R^3_...)$- and
${\cal O}(R^4_...)$-type. Higher order terms have no much
importance, because they do not affect the divergences, in
case of the ``right'' distribution of non-local exponential
factors. Our treatment will cover both polynomial and
exponential choices of the form factors $\Phi=\Phi(\Box)$ and
$\Psi=\Psi(\Box)$ in Eq. (\ref{act Phi}). As we have noted
in the main text, in the exponential case the theory can be
superrenormalizable only if the functions $\Phi$ and $\Psi$
have the same exponential factor, let's call it $\exp(-\al\Box)$.
In the polynomial QG of Ref. \cite{highderi} the requirement
is less rigid, namely $\Phi$ and $\Psi$ should be polynomials
of the same order. For the sake of simplicity, let us assume
that $\Phi=\Psi$. Our consideration will be partially
repeating the one of \cite{Stelle-77,book} and \cite{highderi}
and we include it mainly to provide consistent presentation and
to discuss special features of the non-local case.

We assume that the quantum metric $h_{\mu\nu}$ is defined as
$\,h_{\mu\nu}=g_{\mu\nu}-\eta_{\mu\nu}$.
Let us introduce the gauge fixing condition in the  form
\beq
S_{\rm gf}
&=& \int d^4x \,\, \chi_\mu \,Y^{\mu\nu} \, \chi_\nu\,,
\label{11}
\eeq
with the following form of the gauge-fixing and weight
functions:
\beq
\chi_\mu &=& \pa_\la h_\mu^\la - \be \,\pa_\mu h
\nonumber
\\
Y^{\mu\nu}
&=&
- \frac{1}{\tau} \,\Omega(\Box)\,\left( g^{\mu\nu} \,\Box +
\ga \,\na^\mu\, \na^\nu \right) \,.
\label{12}
\eeq
Here $h=h_{\mu\nu} \eta_{\mu\nu}$  and $\,\be, \,\tau,\,\ga\,$ are
gauge-fixing parameters. Furthermore, $\Om(\Box)$ is a function,
which can be chosen by the convenience criteria.
Our first purpose is to have a non-degenerate bilinear form of
the action, therefore it is useful to choose
$\Om(\Box)=\Phi(\Box)=\Psi(\Box)$. As it was discussed in the main
text, the divergences do not depend on the choice of the gauge
fixing. For this reason we will not discuss
the most general form of the weight function, which may be
dependent on the curvature tensor. Also, the gauge-fixing
parameters can be chosen to make the consideration simpler.

The most general bilinear form of the action on the flat
background is
\beq
S^{(2)} &=&
\frac12\,\int d^4x\,\,h^{\ka\om}
\Big\{
k_1\de_{\ka\om , \rho\si}\,\Box^2
+ k_2g_{\ka\om} g_{\rho\si}\,\Box^2
+ k_3 \big(g_{\ka\om} \pa_\rho\pa_\si
+ g_{\rho\si} \pa_\ka\pa_\om\big)
\nonumber
\\
&+& k_4 \big(g_{\ka\rho} \pa_\om\na_\si
+ g_{\ka\si} \pa_\om\na_\rho
+ g_{\om\rho} \pa_\rho\pa_\si
+ g_{\om\si} \pa_\ka\pa_\rho\big)
+ k_5  \pa_\om\pa_\ka\pa_\rho\pa_\si
\Big\}h^{\rho\si}\,,
\label{biline}
\eeq
Here $\,k_i,\,\,i=1,\dots,5\,$ are some functions of $\Box$, which
depend on the choice of the theory. In case of the polynomial theory
(\ref{superre}) they are polynomials, while in the case of
the exponential theory they are all proportional to
$\,\Box\,\exp\big(-\al\Box \big)$. The explicit form of these
functions can be found in \cite{BGKM} and \cite{Mode-fin}, but we
do not need them here.

Since the gauge-fixing parameters $\,\be$, $\,\tau$
and  $\,\ga\,$ do not affect divergences, one can chose them
in such a way that the bilinear form of the overall action
$\,S^{(2)} + S_{\rm gf}\,$ becomes minimal. This means that the
tensor structures proportional to $\,k_3$, $\,k_4\,$ and $\,k_5\,$
cancel. Then, since the remaining $k_1$ and $k_2$ will be
proportional to $\Phi(\Box)$, the propagator of the
quantum metric has $r_l = 4+2N$ for the $\Phi(\Box)$ being
polynomial of order $N$, and $r_l={\cal I}$ for the exponential.

In order to apply the power counting relations (\ref{dD}) and
(\ref{topo}), one has to provide that the Faddeev-Popov ghosts
have the same power of momenta $r_l$ in the bilinear form of
their action. This can be achieved by using the method suggested
by Fradkin and Tseytlin \cite{frts-82}. The presence of extra
derivatives in the form factors of the initial action does not
affect the scheme \cite{highderi}, in both polynomial and
exponential models. Let us introduce the modified form of the
ghost action,
\beq
S_{gh} &=&
\int d^4x\sqrt{-g}\,
{\bar C}_\al\,Y^\al_{\,\,\be}\,\,M^\be_{\,\,\ga}\,C^\ga\,,
\quad
\mbox{where}
\quad
M^\be_{\,\,\ga} \,=\, \frac{\de \chi^\be}{\de h_{\rho\si}}
\,R_{\rho\si.\ga}\,,
\label{FP}
\eeq
where $R_{\rho\si.\ga}$ is the generator of gauge transformations
of $\,h_{\mu\nu}\,$ in the background field method. An extra
insertion of the weight function in the definition of the ghost
action in (\ref{FP}) provides that for the quantum metric and
for the Faddeev-Popov ghosts there will be the same $r_l$ in the
formula (\ref{dD}).

The effective action $\Ga$ is defined as \cite{frts-82}
\beq
e^{i\Ga[g_{\mu\nu}]}
&=&
\big( \Det Y_\al^\be \big)^{-1/2} \int dh_{\mu\nu} d{\bar C}_\al
dC^\be\, e^{i S + iS_{gf} + iS_{gh}}\,.
\label{Z}
\eeq
The remaining problem is how to evaluate the
functional determinant $\,\Det Y_\al^\be\,$ in the last expression.
For the polynomial QG theory (\ref{superre}), this operator is of
the standard sort,
\beq
Y_\al^\be &=&
\Box^{2k}\,\big(\Box^2\,\de^\al_\be
+ \la \na^\al\na_\be\big)
\,+\,{\cal O}(R_{\dots})\times \na^{2k}_{\dots}\,+\dots\,
\,,
\label{op}
\eeq
considered in \cite{highderi}. This type of operator can be
elaborated by the generalized Schwinger-DeWitt technique of
Barvinsky and Vilkovisky \cite{bavi-85}. The divergent
contribution of this expression is ${\cal O}(R^2_{\dots})$,
confirming the power counting - based analysis.

In the exponential case the contribution of the weight operator
in (\ref{op}) is more complicated. It is easy to see
that $\,\Det Y_\al^\be\,$ is factorized into the product of
determinants of the two operators. One of these operators is
trivial and for the second one has to evaluate
\beq
\log \Det \Om(\Box) &=&
\log \Det \exp \big(-\al\Box\big)
\,=\,
\Tr \log \exp \big(-\al\Box\big)
\,=\,
\Tr \big(-\al\Box\big)\,.
\label{expo}
\eeq
The last expression has quadratic divergences, but it does not mean
there are no logarithmic ones too, as usual. The evaluation of
it can be performed by local momentum representation or by the
Schwinger-DeWitt technique. Unfortunately, the operator
(\ref{expo}) seems to be unappropriate for the
technique of \cite{bavi-85}. However, there are strong reasons
to suppose that the result will be qualitatively the same as for
the polynomial case. In order to see this, one has to regard the
$\,\exp \big(-\al\Box\big)\,$ as a limit of the expression
(\ref{series}). In order to complete
the story, one has to note that a qualitatively different result
for the divergent part of (\ref{expo}) would mean one more
addition to Eq. (\ref{exp-brok}). This would further enforce the
main conclusion of the present paper, concerning the presence
of the infinite set of ghosts in the dressed propagator of the
exponential theory.


\end{document}